\documentclass[preprint,authoryear,12pt]{elsarticle}

\usepackage{color}
\newcommand{\paars}[1]{\textcolor{black}{#1}}

\usepackage{epsfig}
\usepackage{graphicx}
\usepackage{psfrag,pst-grad, pst-plot, pstricks}
\usepackage[latin1]{inputenc} 
\usepackage{caption}
\usepackage{hypernat}

\usepackage{dsfont, amsmath, amssymb}
\usepackage[amsmath, thmmarks]{ntheorem} 

\usepackage{ulem}
\renewcommand{\emph}{\normalem}

\newcommand{\pre}{\theorempreskipamount}
\newcommand{\post}{\theorempostskipamount}
\theoremstyle{break}            
\theoremheaderfont{\normalfont\bfseries}
\theorembodyfont{\itshape}                                                    
\theoremsymbol{\ensuremath{\Box}}
\theoremseparator{\;\;}
\newtheorem{thm}{Theorem}[section]           

\newtheorem*{addendum}{Addendum}

\theoremstyle{plain}
\theoremheaderfont{\normalfont\bfseries}
\theorembodyfont{\itshape}
\theoremsymbol{\ensuremath{\Box}}
\theoremseparator{\;\;}
\newtheorem{prop}[thm]{Proposition}

\theoremstyle{plain}
\theoremheaderfont{\normalfont\bfseries}
\theorembodyfont{\itshape}
\theoremsymbol{\ensuremath{\Box}}
\theoremseparator{\;\;}

\theoremstyle{plain}
\theoremheaderfont{\normalfont\bfseries}
\theorembodyfont{\itshape}
\theoremsymbol{\ensuremath{\Box}}
\theoremseparator{\;\;}
\newtheorem{cor}[thm]{Corollary}

\theoremstyle{plain}
\theoremheaderfont{\scshape}
\theorembodyfont{\normalfont}
\theoremsymbol{\ensuremath{\Box}}
\theoremseparator{\;\;}
\newtheorem{dfn}[thm]{Definition}

\theoremstyle{plain}
\theoremheaderfont{\scshape}
\theorembodyfont{\normalfont}
\theoremsymbol{\ensuremath{\Box}}
\theoremseparator{\;\;}
\theoremnumbering{arabic}
\newtheorem{bsp}[thm]{Example}

\theoremstyle{plain}
\theoremheaderfont{\scshape}
\theorembodyfont{\normalfont}
\theoremsymbol{\ensuremath{\Box}}
\theoremseparator{:\;\;}
\theoremnumbering{arabic}
\newtheorem{rmk}[thm]{Remark}

\theoremstyle{nonumberbreak}
\theoremheaderfont{\scshape}
\theorembodyfont{\normalfont}
\theoremsymbol{\ensuremath{\Box}}
\theoremseparator{:}
\theoremnumbering{arabic}
\newtheorem{rmks}{Remarks}

\theoremstyle{change}
\theoremheaderfont{\sc}
\theorembodyfont{\normalfont}                   % slshape, upshape
\theoremseparator{:\;\;}
\theoremsymbol{\ensuremath{\Box}}  % variante: \rule{1.2ex}{1.2ex} würde eigenes Symbol für proof umgebung kreiren: schwarze box.
\newtheorem*{Proof}{Proof}

\theoremstyle{change}
\theoremheaderfont{\sc}
\theorembodyfont{\normalfont}                   % slshape, upshape
\theoremseparator{:\;\;}
\theoremsymbol{\ensuremath{\Box}}  % variante: \rule{1.2ex}{1.2ex} würde eigenes Symbol für proof umgebung kreiren: schwarze box.

\theoremstyle{change}
\theoremheaderfont{\sc}
\theorembodyfont{\normalfont}
\theoremseparator{:\;\;}
\theoremsymbol{\ensuremath{\Box}}

\newcommand{\be}{\begin{equation}}
\newcommand{\ee}{\end{equation}}
\newcommand{\bd}{\begin{displaymath}}
\newcommand{\ed}{\end{displaymath}}
\newcommand{\md}{\mathds}

\newcommand{\bc}{\begin{center}}
\newcommand{\ec}{\end{center}}

\title{Risk Concentration and Diversification:\\Second-Order Properties}

\begin{document}

\author[md]{Matthias Degen\corref{cor1}}
\ead{degen@math.ethz.ch} 
\address[md]{Department of Mathematics, ETH Zurich, 8092 Zurich, Switzerland}
\cortext[cor1]{Corresponding author. Tel.: +41 44 632 34 28} 

\author[ddl]{Dominik D.~Lambrigger} 
\address[ddl]{Department of Mathematics, ETH Zurich, 8092 Zurich, Switzerland}

\author[js]{Johan Segers} 
\address[js]{Institut de statistique, Universit\'e catholique de Louvain, B-1348 Louvain-la-Neuve, Belgium}

\begin{frontmatter}

\begin{abstract}
The quantification of diversification benefits due to risk aggregation plays a prominent role in the (regulatory) capital management of large firms within the financial industry. However, the complexity of today's risk landscape makes a quantifiable reduction of risk concentration a challenging task. In the present paper we discuss some of the issues that may arise. The theory of second-order regular variation and second-order subexponentiality provides the ideal methodological framework to derive second-order approximations for the risk concentration and the diversification benefit.
\end{abstract}


\end{frontmatter}

%%%%%%%%%%%%%%%%%%%%%%%%%%%%%%%%%%%%%%%%%%%%%%%%%%%%
%%%%%%%%%%%%%%%%%%%%%%%%%%%%%%%%%%%%%%%%%%%%%%%%%%%%
%%%%%%%%%%%%%%%%%%%%%%%%%%%%%%%%%%%%%%%%%%%%%%%%%%%%
%%%%%%%%%%%%%%%%%%%%%%%%%%%%%%%%%%%%%%%%%%%%%%%%%%%%

\section{Introduction}

Diversification is one of the most popular techniques to mitigate exposure to risk and constitutes an important part within the current regulatory framework for banks (Basel II) as well as in the preparation of the new regulatory framework for insurance companies (Solvency II). However, due to the increasing complexity of financial and insurance products, a quantitative analysis of diversification benefits has become a demanding task. Many authors have warned against an imprudent application of diversification concepts, especially when the underlying risk factors show a heavy-tailed pattern; see for instance \cite{FAMA}, p.~269, \cite{ROOTZEN}, \cite{EMS} or \cite{DEGEN}.  More recently, \cite{IW07} and \cite{IJW09} have discussed diversification benefits linking heavy-tailed distributions to specific economic models.

In addition to that, the structure of internationally active financial groups has created the need to analyze diversification effects not only at the individual firm or subsidiary level, but also at the group level. For recent advances in the context of group supervision, see for instance \cite{Gatzert} or \cite{Gatzert-al} and references therein.

In the present paper we study mathematical properties of diversification effects under the risk measure Value-at-Risk (VaR). For a risky position $X$ with distribution function $F$, the {\it Value-at-Risk} at the level $\alpha$ is defined by VaR$_\alpha(X) = F^{\leftarrow}(\alpha)$, $0 < \alpha < 1$, where $F^{\leftarrow}(\alpha) = \inf\left\{x \in \md{R}: F(x) \geq \alpha\right\}$, denotes the generalized inverse of $F$. Under the Basel II/Solvency II framework, VaR$_\alpha(X)$ essentially corresponds to the regulatory risk capital a financial institution needs to hold in order to be allowed to carry the risky position $X$ on its books. Note that the level $\alpha$ is given by the respective regulatory authority and is typically close to 1. 

Throughout we assume that our potential future losses $X_1,\ldots,X_n$, $n \geq 2$, are non-negative independent and identically distributed (iid) random variables with continuous distribution function $F$. We write $\overline{F} = 1-F$ for the tail of $F$. We assume that $\overline{F} \in RV_{-1/\xi}$, i.e.~for every $x > 0$,
\bd
\frac{\overline{F}(tx)}{\overline{F}(t)} \rightarrow x^{-1/\xi}, \quad t \rightarrow \infty,
\ed
so that
\be\label{first-order-divben}
C(\alpha) = \frac{\textnormal{VaR}_{\alpha}\left(\sum_{k =1}^n X_k\right)}{\sum_{k = 1} ^n \textnormal{VaR}_{\alpha}(X_k)}\;\; \rightarrow\;\; n^{\xi -1}, \quad \alpha\rightarrow 1.
\ee
We will refer to $C(\alpha)$ as the {\it risk concentration} (at level $\alpha$) and to $1 - C(\alpha)$ as the {\it diversification benefit}. 

For our mathematical discussion below, we do not restrict ourselves to a specific economic interpretation of the rvs $X_k$, but---with regard to possible applications---the reader might want to think of the following examples. Within the Basel II framework, the $X_k$'s could for instance represent the yearly aggregated operational risk (OR) losses of a business line $k$ on an individual firm or subsidiary level; see \cite{BIS2006}, \textsection\;657, 669. Or, in the context of group supervision, the $X_k$'s might be viewed as the total yearly OR loss of a subsidiary company $k$; see \cite{BIS2006}, \textsection\;657. In a Solvency II framework, one might think of the $X_k$'s as the assets and/or liabilities of a subsidiary company $k$; see for instance \cite{Gatzert-al} for a discussion of risk concentration results in the non-heavy-tailed case.

Due to the non-coherence of VaR, diversification benefits may be positive or negative. A risk concentration value $C(\alpha)$ greater than 1 means non-diversification (i.e.~superadditivity of VaR$_{\alpha}$) at the level $\alpha$. In such cases, aggregation of risks would even lead to an increase of regulatory risk capital. 

On a side note we remark that the issue of non-diversification does not arise for instance in the Swiss Solvency Test (SST), as there the coherent risk measure Expected Shortfall (ES) is used; see \cite{ADEH99}. Nevertheless, the asymptotic result (\ref{first-order-divben}) still holds under ES (provided that $\xi < 1$).\\

\noindent
Given the high $\alpha$-levels typically of interest for risk management practice, analyzing risk concentration by means of its empirical counterpart will in general not yield much insight. One is therefore advised to consider (suitable) analytic approximations of $C(\cdot)$. In the simple case of $n$ regularly varying iid losses, \paars{relation}~(\ref{first-order-divben}) gives rise to a first-order approximation $C_1(\alpha) \equiv n^{\xi -1}$ of $C(\alpha)$ for $\alpha$ close to 1. \paars{The} asymptotic result (\ref{first-order-divben}) has been generalized to situations where the vector ${\bf X} = (X_1, \ldots, X_n) \geq {\bf 0}$ is multivariate regularly varying with index $-1/\xi$ and with identically distributed margins; see \cite{Barbe-Foug-Gen}, Proposition 2.1 and \cite{ELW}, Theorem 4.1.

The main issue discussed in this paper is that the convergence in (\ref{first-order-divben}) may be arbitrarily slow. As a consequence, in risk management practice, where we are interested in $C(\alpha)$ at some fixed level of $\alpha$ typically ranging from $95\%$ to $99.97\%$, the first-order approximation $C_1(\alpha)$ may be too crude as the following example illustrates.

\begin{figure}[htb!]
\begin{center}
\setcaptionwidth{\textwidth}
   \epsfig{file=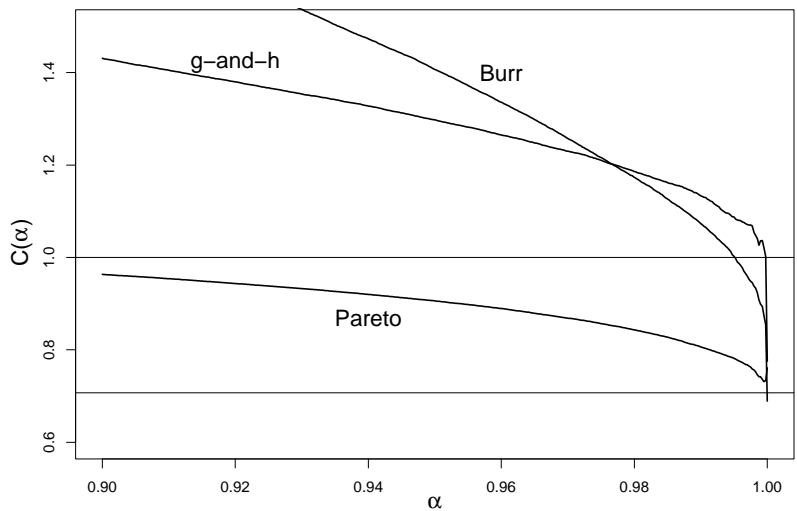,width=0.6\textwidth, angle=-90}
   \caption{Empirical risk concentration (based on $10^7$ simulations) together with a first-order approximation $C_1 \equiv 1/\sqrt{2}\approx 0.71$ for two iid random variables from a Burr ($\tau = 0.25, \kappa = 8$), a Pareto ($\xi = 0.5$) and a g-and-h ($a=0,b=1,g=2, h=0.5$) distribution; see Remark \ref{rmk:pareto} and Examples \ref{ex:Burr} and \ref{ex:gandh} for the parameterization used in the respective models.}
   \label{fig1}
  \end{center}
\end{figure}

\begin{bsp}\label{bsp1}
Consider a financial institution holding two risks $X_1$ and $X_2$. Assume that these positions are modeled by $X_1, X_2 \stackrel{iid}{\sim} F$ with $\overline{F} \in RV_{-2}$. In that case the risk concentration satisfies $C(\alpha) \rightarrow 2^{\xi -1} = 1/\sqrt{2} \approx 0.71$, for $\alpha \rightarrow 1$. Therefore, when aggregating $X_1$ and $X_2$, a diversification benefit (reduction of regulatory risk capital) of about 29\% would seem reasonable at first sight (for high levels of $\alpha$). Figure \ref{fig1} however shows that such an argumentation needs careful rethinking. 
\end{bsp}

\noindent
In what follows, we suggest the reader to keep Figure~\ref{fig1} in mind as \paars{a warning against a careless use of asymptotics to justify diversification benefits}. Most importantly, the behavior of the risk concentration $C(\alpha)$ at levels of $\alpha$ close to 1 (typically of interest for applications) may be very sensitive, i.e.~small changes of $\alpha$ \paars{may} lead to large changes of $C(\alpha)$. In economic terms this means that while we may well expect diversification benefits of considerable size at a certain level $\alpha$, this may change rather drastically into non-diversification once we move away only little from that level.

Altogether this motivates the consideration of a second-order approximation for the risk concentration $C$. Concerning methodology, we draw on the theories of {\it second-order regular variation} and {\it second-order subexponentiality}. Our main result, Theorem~\ref{thm1}, gives the precise asymptotic behavior of \paars{the approximation error} $C(\alpha) - n^{\xi-1}$ as $\alpha \rightarrow 1$. As it turns out, asymptotically two situations may arise. Without going into details at this point, it is either the asymptotic behavior in the second-order regular variation part \paars{or the one} in the second-order subexponential part that dominates in the limit. 

For the more applied risk management end-user, the main message is that, not only for infinite mean but also for finite mean \paars{models used} in financial and insurance risk management, aggregating risks may somewhat surprisingly result in a negative diversification benefit (under VaR).

The paper is organized as follows. Section~2 recalls basic definitions and results on second-order regular variation and second-order subexponentiality. In Section~3 we present our main result on the second-order behavior of the risk concentration $C(\alpha)$. In Section~4 we apply this result to different distribution functions relevant for practice.

%%%%%%%%%%%%%%%%%%%%%%%%%%%%%%%%%%%%%%%%%%%%%%%%%%%%
%%%%%%%%%%%%%%%%%%%%%%%%%%%%%%%%%%%%%%%%%%%%%%%%%%%%
%%%%%%%%%%%%%%%%%%%%%%%%%%%%%%%%%%%%%%%%%%%%%%%%%%%%
%%%%%%%%%%%%%%%%%%%%%%%%%%%%%%%%%%%%%%%%%%%%%%%%%%%%

\section{Preliminaries}
The tail quantile function associated with the distribution function $F$ is denoted by $U_F(t) = (1/\overline{F})^{\leftarrow}(t) = F^{\leftarrow}(1-1/t)$, $t > 1$. Where clear from the context, we omit the subscript and write $U$ instead of $U_F$. Recall that $\overline{F} \in RV_{-1/\xi}$ for some $\xi > 0$ is equivalent to $U \in RV_{\xi}$. In this case we write $U(t) = t^{\xi} L_U(t)$, where $L_U \in RV_0$ denotes the slowly varying function associated with $U$. 

The distribution function of the sum $X_1 + \cdots + X_n$ is denoted by $G$ and due to the iid assumption we have $G(x) = F^{n\ast}(x)$, the $n$-fold convolution of $F$. Since $\overline{F}$ is regularly varying, \paars{$F$} is subexponential and hence
\bd
\frac{\overline{G}(x)}{\overline{F}(x)} \rightarrow n, \quad x \rightarrow \infty;
\ed
see for instance \cite{EKM}, Corollary 1.3.2. In terms of quantiles, setting $G^{\leftarrow}(\alpha) = x$, we obtain
\be\label{first-order-quantiles}
\frac{G^{\leftarrow}(\alpha)}{F^{\leftarrow}(\alpha)} = \frac{U_F\left(1/\overline{F}(x)\right)}{U_F\left(1/\overline{G}(x)\right)} =  \frac{U_F\left(1/\overline{F}(x)\right)}{U_F\big(\underbrace{\overline{F}(x)/\overline{G}(x)}_{\rightarrow 1/n}1/\overline{F}(x)\big)} \rightarrow n^{\xi}, \quad \alpha \rightarrow 1,
\ee
due to the Uniform Convergence Theorem for regularly varying functions; see for instance \cite{BINGHAM}, Theorem 1.5.2. This implies
\bd
C(\alpha) = \frac{\textnormal{VaR}_{\alpha}\left(\sum_{k =1}^n X_k\right)}{\sum_{k = 1} ^n \textnormal{VaR}_{\alpha}(X_k)}\;\; \rightarrow\;\; n^{\xi -1}, \quad \alpha\rightarrow 1.
\ed

\noindent
In order to analyze the convergence rate of $C(\alpha)$ to $n^{\xi-1}$ as $\alpha \rightarrow 1$, \paars{the derivation in} (\ref{first-order-quantiles}) suggests to study the second-order behavior \paars{in the two limit relations}
\begin{align}
\label{RV-U}
\frac{U(ts)}{U(t)} &\rightarrow s^{\xi} , \quad t \rightarrow \infty,\\
\label{subexp}
\frac{\overline{G}(x)}{\overline{F}(x)} &\rightarrow n, \quad x \rightarrow \infty.
\end{align}

\noindent
Rate of convergence results for (\ref{RV-U}) are well-established within the framework of {\it second-order regular variation}; see for instance \cite{deHaan}, Section 2.3 and Appendix B.3 for an introduction. Rate of convergence results for (\ref{subexp}) may be obtained using the framework of {\it second-order subexponentiality}; see for instance \cite{Omey, Omey2} and \cite{Barbe}. Below we review these two concepts.

%%%%%%%%%%%%%%%%%%%%%%%%%%%%%%%%%%%%%%%%%%%%%%%%%%%%
%%%%%%%%%%%%%%%%%%%%%%%%%%%%%%%%%%%%%%%%%%%%%%%%%%%%

\subsection{Second-order regular variation}
\begin{dfn}[Second-order regular variation]\label{dfn:2RV}
\paars{A function} $U \in RV_{\xi}$ with $\xi > 0$ is said to be of second-order regular variation with parameter $\rho \leq 0$, if there exists a function $a(\cdot)$ with $\displaystyle{\lim_{t\rightarrow \infty}a(t) = 0}$ such that
\be
\label{2RV}
\lim_{t\rightarrow \infty}\frac{\frac{U(ts)}{U(t)} - s^{\xi}}{a(t)} = H_{\xi,\rho}(s)=s^\xi \frac{s^\rho-1}{\rho},
\ee
with the obvious interpretation for $\rho = 0$. In this case we write $U \in 2RV_{\xi, \rho}(a)$ and refer to $a(\cdot)$ as the {\it auxiliary function} of $U$. 
\end{dfn}

\noindent
Note that $U \in 2RV_{\xi,\rho}$ is equivalent to $\overline{F} \in 2RV_{-1/\xi, \rho/\xi}$ for $\xi > 0, \rho \leq 0$. It is well known that if a non-trivial limit $H_{\xi,\rho}$ in (\ref{2RV}) exists which is not a multiple of $s^\xi$, then it is necessarily of the form stated. Furthermore, the auxiliary function satisfies $|a| \in RV_{\rho}$; see for instance \cite{HAANSTADTMUELLER}, Theorem 1. The second-order parameter $\rho$ thus governs the rate of convergence in (\ref{RV-U}), i.e.~the smaller $|\rho|$, the slower the convergence.

A broad and frequently used subclass of models satisfying (\ref{2RV}) is given by the so-called Hall class; see also \cite{hall}.

\begin{dfn}[Hall Class]
A distribution function $F$ is said to belong to the {\it Hall class} if its quantile function $U$ admits the asymptotic representation \paars{$U(t) = c \, t^{\xi}\bigl(1 + d\, t^{\rho} + o(t^{\rho})\bigr)$} as $t \rightarrow \infty$, for some $c>0$, $d \in \mathds{R}\setminus\{0\}$, and first- and second-order parameters $\xi > 0$ and $\rho<0$. 
\end{dfn}

\noindent
In terms of tail functions this means that we consider models of the form
\bd
\overline{F}(x) = \left(\frac{x}{c}\right)^{-1/\xi}\Big( 1 + \frac{d}{\xi}\left(\frac{x}{c}\right)^{\rho/\xi} +o\big(x^{\rho/\xi}\big)\Big), \quad x \rightarrow \infty.
\ed
Note that the tail quantile function of a loss model in the Hall class obviously satisfies $U \in 2RV_{\xi, \rho}(a)$ with $a(t) \sim d\rho t^{\rho}$ as $t \rightarrow \infty$. \paars{[Throughout the paper we mean by $f_1(t) \sim f_2(t)$ for $t \rightarrow t_0$ that $f_1(t)/f_2(t) \to 1$ as $t \rightarrow t_0$.]} Conversely, loss models that are second-order regularly varying with $\rho < 0$ and with auxiliary function $a(t) \sim d \rho t^\rho$ for $t \rightarrow \infty$ are members of the Hall class. This follows from the Representation Theorem for extended regularly varying functions; see \cite{BINGHAM}, Theorem 3.6.6.

\begin{rmk}
\label{rmk:pareto}
For the standard Pareto model $U(t) = t^{\xi}$, the convergence in (\ref{RV-U}) is immediate. We interpret this case as $U \in 2RV_{\xi,-\infty}$. 
\end{rmk}

\noindent
Heavy-tailed models $U \in 2RV_{\xi,\rho}$ not belonging to the Hall class include for instance the loggamma or the g-and-h distribution ($\rho = 0$ in both cases).

%%%%%%%%%%%%%%%%%%%%%%%%%%%%%%%%%%%%%%%%%%%%%%%%%%%%
%%%%%%%%%%%%%%%%%%%%%%%%%%%%%%%%%%%%%%%%%%%%%%%%%%%%

\subsection{Second-order subexponentiality}
Second-order subexponentiality results by \cite{Barbe} (Theorems 2.2 and 2.5) are summarized in the following proposition; see also \cite{Omey, Omey2} for similar results.

\begin{prop}\label{prop1}\NoEndMark
Assume that $F$ is differentiable with $F(0)=0$. If for some $\xi > 0$, $\overline{F}(x)/\left(x \, f(x)\right) \rightarrow \xi$ as $x \rightarrow \infty$, then for $n \geq 2$ and with $G(x) = F^{n\ast}(x)$,

\bd
\lim_{x \rightarrow \infty}\frac{\frac{\overline{G}(x)}{\overline{F}(x)} - n}{b(x)} = J_{\xi}(n) = n(n-1)c_\xi,
\ed
with
\bd
c_\xi = \begin{cases}
1/\xi, & \textrm{if}\; \xi \leq 1,\\
(1-\xi)\frac{\Gamma^2(1-1/\xi)}{2\Gamma(1-2/\xi)}, & \textrm{if}\; \xi > 1, \\
\end{cases} 
\ed 
and
\bd
b(x) = \begin{cases}
\mu_F/x, &\textrm{if}\; \xi \leq 1, \mu_F < \infty,\\
\mu_F(x)/x, &\textrm{if}\; \xi = 1, \mu_F = \infty,\\
\overline{F}(x)/(\xi-1), &\textrm{if}\; \xi > 1,
\end{cases}
\ed

\noindent
where $\mu_F(x) = \int_0^{x}t\,dF(t)$, $\mu_F = \displaystyle{\lim_{x\rightarrow \infty}}\mu_F(x)$ and where $\Gamma$ denotes the gamma function.
\end{prop}

\begin{figure}[htb!]
\begin{center}
\setcaptionwidth{11cm}
   \epsfig{file=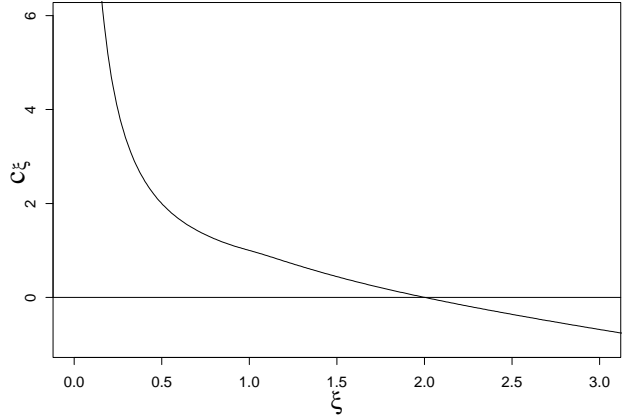,width=6.5cm, angle=-90}
   \caption{$c_\xi$ in Proposition \ref{prop1} as a function of $\xi$.}
   \label{fig:xi}
  \end{center}
\end{figure}

\begin{rmks}\NoEndMark
\begin{itemize}
\item[i)]
In defining $c_\xi$ \paars{for $1 < \xi < 2$} we make use of the analytic continuation of the gamma function $\Gamma$ to $\mathds{C}\backslash\{0,-1, -2,\ldots\}$; see \cite{ABR-STEGUN}, Formula~6.1.2. 
\item[ii)]  Note that $c_\xi$ is strictly decreasing in $\xi$ and thus $c_\xi = 0$ if and only if $\xi = 2$; see also Figure \ref{fig:xi}. In that case, Proposition \ref{prop1} does not yield a (proper) second-order result for convolutions in that particular case. To the best of our knowledge, second-order asymptotics of the above form for $\xi=2$ are not available in the literature (except for special cases, such as stable laws).
\item[iii)] In the case $\xi = 1$ and $\mu_F = \infty$, we have $f = F' \in RV_{-2}$. Karamata's Theorem implies that $\mu_F(x) = \int_{0}^{x} t f(t) dt$ is slowly varying; see for example \cite{deHaan}, Theorem B.1.5. 
\item[iv)] By Proposition \ref{prop1}, the asymptotic behavior of the function $b$ is fully specified and we have $b \in RV_{-(1\wedge 1/\xi)}$ with $\xi > 0$.
\end{itemize}
\end{rmks}

%%%%%%%%%%%%%%%%%%%%%%%%%%%%%%%%%%%%%%%%%%%%%%%%%%%%
%%%%%%%%%%%%%%%%%%%%%%%%%%%%%%%%%%%%%%%%%%%%%%%%%%%%
%%%%%%%%%%%%%%%%%%%%%%%%%%%%%%%%%%%%%%%%%%%%%%%%%%%%
%%%%%%%%%%%%%%%%%%%%%%%%%%%%%%%%%%%%%%%%%%%%%%%%%%%%

\section{Main result}
Combining the concepts of second-order regular variation and second-order subexponentiality, we obtain a second-order result for the risk concentration $C$. It may be viewed as a partial quantile analogue of Theorem 3.2 in \cite{Geluk}. \paars{Recall the notations in the previous sections, in particular in Proposition~\ref{prop1}.}

\begin{thm}\label{thm1}\NoEndMark

Let $X_1,\ldots,X_n \stackrel{iid}{\sim} F$ be positive random variables and let $U = (1/\overline{F})^{\leftarrow}$ be such that $t\,U'(t)/U(t) \rightarrow \xi > 0$. Suppose \paars{that} $U \in 2RV_{\xi, \rho}(a)$ for some $\rho \leq 0$ and \paars{with auxiliary} function $a(\cdot)$ of ultimately constant sign. \paars{If $\rho \neq -(1 \wedge \xi)$}, then, for \paars{fixed} $n \geq 2$ and \paars{as} $\alpha \rightarrow 1$,
\bd
C(\alpha) = \frac{\textnormal{VaR}_{\alpha}\left(\sum_{k =1}^n X_k\right)}{\sum_{k = 1} ^n \textnormal{VaR}_{\alpha}(X_k)}
= n^{\xi-1} + K_{\xi,\rho}(n) A(\alpha) + o\bigl(A(\alpha)\bigr)
\ed
\paars{where $A(\alpha)$ and $K_{\xi,\rho}(n)$ are given as follows:}
\begin{itemize}
\item[(i)] \paars{case $\rho < -(1 \wedge \xi)$:}
\begin{align*}
\hspace{-1cm}  A(\alpha) = b\bigl(F^\leftarrow(\alpha)\bigr) &=
  \begin{cases}
    \mu_F / F^\leftarrow(\alpha), & \text{if $\xi \le 1$, \paars{$\rho < -\xi$}, $\mu_F < \infty$,} \\
    \mu_F\bigl(F^\leftarrow(\alpha)\bigr) / F^\leftarrow(\alpha), & \text{if $\xi = 1$, $\rho < -1$, $\mu_F = \infty$,} \\
    (1-\alpha) / \paars{(\xi - 1)}, & \text{if $\xi > 1$, $\rho < -1$;}
  \end{cases} \\
  K_{\xi,\rho}(n) &=
  \begin{cases}
    \paars{(n-1)/n}, & \text{if $\xi \le 1$, $\rho < -\xi$,} \\
    \paars{n^{\xi-2}(n-1) \, \xi \, c_\xi}, & \text{if $\xi > 1$, $\rho < -1$;}
  \end{cases}
\end{align*}
\item[(ii)] \paars{case $\rho > -(1 \wedge \xi)$:}
\begin{align*}
  A(\alpha) &= a\bigl(1/(1-\alpha)\bigr), \\
  K_{\xi,\rho}(n) &= n^{\xi-1} \, \frac{n^\rho-1}{\rho}.
\end{align*}
\end{itemize}
\end{thm}

\begin{Proof}
See Appendix.
\end{Proof}

\noindent
As an approximation to the risk concentration $C(\alpha)$, Theorem \ref{thm1} suggests to consider a second-order approximation $C_2(\alpha) = n^{\xi-1} + K_{\xi,\rho}(n) A(\alpha)$, for $\alpha < 1$. 

\begin{figure}[!ht]
\setcaptionwidth{11cm}
\begin{center}

\begin{minipage}[t]{.5\textwidth}
{\LARGE
\psset{unit=2cm}	
\begin{pspicture}(1,-1)(0,0)
\pscustom[linewidth=0pt,linecolor=white,liftpen=0]{
\pspolygon[linewidth=0pt](0,0)(1,-1)(3.25,-1)(3.25,0)  
\fill[fillstyle=hlines,hatchcolor=gray,hatchwidth=0.5pt,hatchsep=2pt]
}
\pscustom[linewidth=0pt,linecolor=white,liftpen=5]{
\pspolygon[linewidth=0pt](0,0)(1,-1)(3.25,-1)(3.25,-1)(3.25,-2.25)(0,-2.25)(0,0)  
\fill[fillstyle=vlines,hatchcolor=gray,hatchwidth=0.5pt,hatchsep=4pt] 
} 

\psline(0,0)(1,-1)(3.25,-1)
\psaxes[showorigin=false]{-}(0,0)(3.5,0.25)
\psaxes[showorigin=false]{-}(0,0)(0,0)(-0.25,-2.4)
  \rput(1.7,0.3){{\fontsize{15}{70}\selectfont  $\mathbf{\xi}$}}
  \rput(-0.3,-1.3){{\fontsize{15}{70}\selectfont  $\mathbf{\rho}$}}
      \rput(2.1,-0.55){{\Large \bf $\rho > -(1\wedge \xi)$}}
         \rput(2.1,-.85){{\fontsize{11}{70}\selectfont \paars{slow} convergence }}
        \rput(1.8,-1.6){{\Large \bf $\rho < -(1\wedge \xi)$ }}
         \rput(1.8,-1.9){{\fontsize{11}{70}\selectfont \paars{fast} convergence }}
\end{pspicture}
}
 \end{minipage}
\end{center}
\vspace{2cm}
 \caption{Illustration of the three possible cases, $\rho > -(1 \wedge \xi)$ (fast convergence), $\rho < -(1 \wedge \xi)$ (slow convergence) and $\rho = -(1 \wedge \xi)$ (boundary case) in the $(\xi, \rho)$--parameter space for loss models $U \in 2RV_{\xi,\rho}$.}
 \label{fig:gamma-rho-diag}
\end{figure}
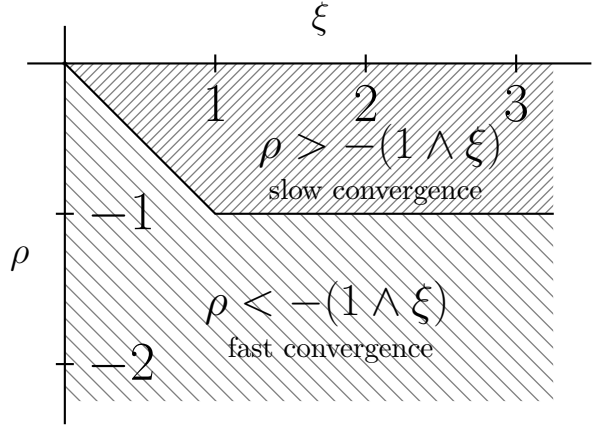

According to Theorem \ref{thm1} two situations arise. Note that
\bd
A(\alpha) = \begin{cases}
b\bigl(F^{\leftarrow}(\alpha)\bigr), &\textrm{if}\;  \rho < -(1 \wedge \xi),\\
a\bigl(1/(1-\alpha)\bigr), &\textrm{if}\; \rho > -(1 \wedge \xi),
\end{cases}
\ed
where $b\circ F^{\leftarrow} \in RV_{-(1 \wedge \xi)}$ and $|a| \in RV_\rho$. \paars{Now if $\rho < -(1 \wedge \xi)$, then $b\bigl(F^{\leftarrow}(\alpha)\bigr)$ vanishes faster than $a\bigl(1/(1-\alpha)\bigr)$ as $\alpha \rightarrow 1$. This motivates the following terminology: a loss model is said to be} of {\it fast convergence} if $U \in 2RV_{\xi,\rho}$ with first- and second-order parameters satisfying $\rho < -(1 \wedge \xi)$, and of {\it slow convergence}, if $\rho > -(1 \wedge \xi)$; see Figure \ref{fig:gamma-rho-diag}.

%%%%%%%%%%%%%%%%%%%%%%%%%%%%%%%%%%%%%%%%%%%%%%%%%%%%
%%%%%%%%%%%%%%%%%%%%%%%%%%%%%%%%%%%%%%%%%%%%%%%%%%%%
%%%%%%%%%%%%%%%%%%%%%%%%%%%%%%%%%%%%%%%%%%%%%%%%%%%%
%%%%%%%%%%%%%%%%%%%%%%%%%%%%%%%%%%%%%%%%%%%%%%%%%%%%

\vspace{1cm}

%%%%%%%%%%%%%%%%%%%%%%%%%%%%%%%%%%%%%%%%%%%%%%%%%%%%
%%%%%%%%%%%%%%%%%%%%%%%%%%%%%%%%%%%%%%%%%%%%%%%%%%%%
%%%%%%%%%%%%%%%%%%%%%%%%%%%%%%%%%%%%%%%%%%%%%%%%%%%%
%%%%%%%%%%%%%%%%%%%%%%%%%%%%%%%%%%%%%%%%%%%%%%%%%%%%

\noindent
\paars{A refinement of the proof of Theorem~\ref{thm1}} also allows to treat the boundary case $\rho =  -(1 \wedge \xi)$. This case seems to be of less relevance for practical applications though.

\begin{addendum}\NoEndMark
Let $U \in 2RV_{\xi,\rho}(a)$ satisfy the conditions of Theorem \ref{thm1}. Suppose, in addition, that for some $q \in \mathds{R}\setminus\{0\}$
\bd
\frac{b(F^{\leftarrow}(\alpha))}{a\left(\frac{1}{1-\alpha}\right)} \rightarrow q, \quad \alpha \rightarrow 1,
\ed
where $b(\cdot)$ is as in Proposition \ref{prop1}. Then, for $n \geq 2$ and $\alpha \rightarrow 1$,

\bd
C(\alpha) = n^{\xi-1} + \left(\xi n^{\xi-2}n^{-(1 \wedge \xi)}J_{\xi}(n)q + n^{-1}H_{\xi,\rho}(n)\right)a\left(\frac{1}{1-\alpha}\right) + o\left(a\left(\frac{1}{1-\alpha}\right)\right),
\ed
with $H_{\xi,\rho}(\cdot)$ and $J_{\xi}(\cdot)$ as in Definition \ref{dfn:2RV} and Proposition \ref{prop1} respectively.
\end{addendum}

\begin{rmk}\label{rmk:aux-function}\NoEndMark
Assume that $U \in 2RV_{\xi, \rho}(a)$. If the associated slowly varying function $L_U$ is differentiable with ultimately monotone derivative $L'_U$, then the auxiliary function $a(\cdot)$ can be chosen as
\bd
a(t) = \frac{t \, U'(t)}{U(t)} - \xi.
\ed 
A proof is given in the Appendix.
\end{rmk}

%%%%%%%%%%%%%%%%%%%%%%%%%%%%%%%%%%%%%%%%%%%%%%%%%%%%
%%%%%%%%%%%%%%%%%%%%%%%%%%%%%%%%%%%%%%%%%%%%%%%%%%%%
%%%%%%%%%%%%%%%%%%%%%%%%%%%%%%%%%%%%%%%%%%%%%%%%%%%%
%%%%%%%%%%%%%%%%%%%%%%%%%%%%%%%%%%%%%%%%%%%%%%%%%%%%

\section{Examples}
In this section we consider the situation of Theorem \ref{thm1} for different loss models. For notational convenience we focus on the case $\rho \neq -(1 \wedge \xi)$. For the Hall class, \paars{Theorem~\ref{thm1} specializes} as follows.

\begin{cor}\label{cor-hall}\NoEndMark
Let $U$ belong to the Hall class \paars{(i.e.\ $U(t) = c\,t^{\xi}\bigl(1 + d\,t^{\rho} + o(t^{\rho})\bigr)$ as $t \rightarrow \infty$, for some $c>0$, $d \in \mathds{R}\setminus\{0\}$, $\xi > 0$, and $\rho<0$)} and satisfy the assumptions of Theorem~\ref{thm1}. Then, the function $A(\cdot)$ in Theorem~\ref{thm1} can be chosen as
\bd
A(\alpha) = \begin{cases}
\paars{\frac{\mu_F}{c}\left(1 - \alpha \right)^{\xi}}
, &\textrm{if}\;  \xi \leq 1,\rho < -\xi, \mu_F<\infty,\\
\paars{-(1- \alpha)\log{(1-\alpha)}}
, &\textrm{if}\;  \xi = 1,\rho < -1, \mu_F=\infty,\\
\paars{(1-\alpha)/(\xi-1)}, &\textrm{if}\;  \xi > 1,\rho < -1,\\
\paars{d\rho(1-\alpha)^{-\rho}}
, &\textrm{if}\; \rho > -(1 \wedge \xi).
\end{cases}
\ed
\end{cor}

\begin{bsp}[Burr]\label{ex:Burr}

Let $X_1, \ldots, X_n \stackrel{iid}{\sim} \textnormal{Burr}(\tau,\kappa)$, with tail function $\overline{F}(x) = (1 + x^{\tau})^{-\kappa}$ for some $\tau,\kappa > 0$. In terms of its tail quantile function this writes as
\be\label{burr}
U(t) =\big(t^{1/\kappa} - 1\big)^{1/\tau} = t^{1/\paars{(\tau\kappa)}}\Big(1 - \frac{1}{\tau}t^{-1/\kappa} + o\big(t^{-1/\kappa}\big)\Big), \quad t \rightarrow \infty,
\ee
so that $U$ belongs to the Hall class with parameters \paars{$c=1$, $d = -1/\tau$, $\xi = 1/\paars{(\tau\kappa)}$, and $\rho = -1/\kappa$}. By (\ref{burr}) the tail quantile function $U$ is given in an explicit form as well as through an asymptotic expansion, so that we may use either Theorem~\ref{thm1} or Corollary~\ref{cor-hall} to derive a second-order result for $C$. Using the former we obtain in the case of fast convergence, i.e.~for \paars{$\kappa \wedge (1/\tau) < 1$},
\bd
C(\alpha) =  \begin{cases}
 n^{ 1/\paars{(\tau\kappa)} -1} + \paars{\Bigl( \frac{n-1}{n} \kappa \, B(\kappa-\frac{1}{\tau}, 1 + \frac{1}{\tau}) + o(1) \Bigr) \, (1-\alpha)^{1/\paars{(\tau\kappa)}}},
&\textrm{if}\; \tau\kappa > 1,\\
1 - \paars{\bigl(\frac{n-1}{n}+o(1)\bigr)(1-\alpha)\log{(1-\alpha)}},
&\textrm{if}\; \tau\kappa = 1,\\
n^{ 1/\paars{(\tau\kappa)} -1} - \frac{n-1}{\tau\kappa}n^{1/\paars{(\tau\kappa)}-2}\frac{\Gamma^2(1-\tau\kappa)}{2\Gamma(1-2\tau\kappa)}(1-\alpha) + o\left(1-\alpha\right), &\textrm{if}\; \tau\kappa < 1,\\
\end{cases}
\ed
 as $\alpha \rightarrow 1$ and where $B(x,y) = \frac{\Gamma(x)\Gamma(y)}{\Gamma(x+y)}$ denotes the \paars{Beta} function. 

In case of slow convergence, i.e.~if \paars{$\kappa \wedge (1/\tau) > 1$}, Theorem~\ref{thm1} together with Remark~\ref{rmk:aux-function} suggests to consider the expansion
\bd
C(\alpha) = n^{1/\paars{(\tau\kappa)} - 1} +
\paars{\frac{1}{\tau} n^{1/\paars{(\tau\kappa)} - 1} (1- n^{-1/\kappa}) (1-\alpha)^{1/\kappa}}
+ o((1-\alpha)^{1/\kappa}), 
\quad \alpha \rightarrow 1.
\ed
\end{bsp}

\begin{figure}[!htb]
\setcaptionwidth{\textwidth}
\begin{center}
      \epsfig{angle=-90,file=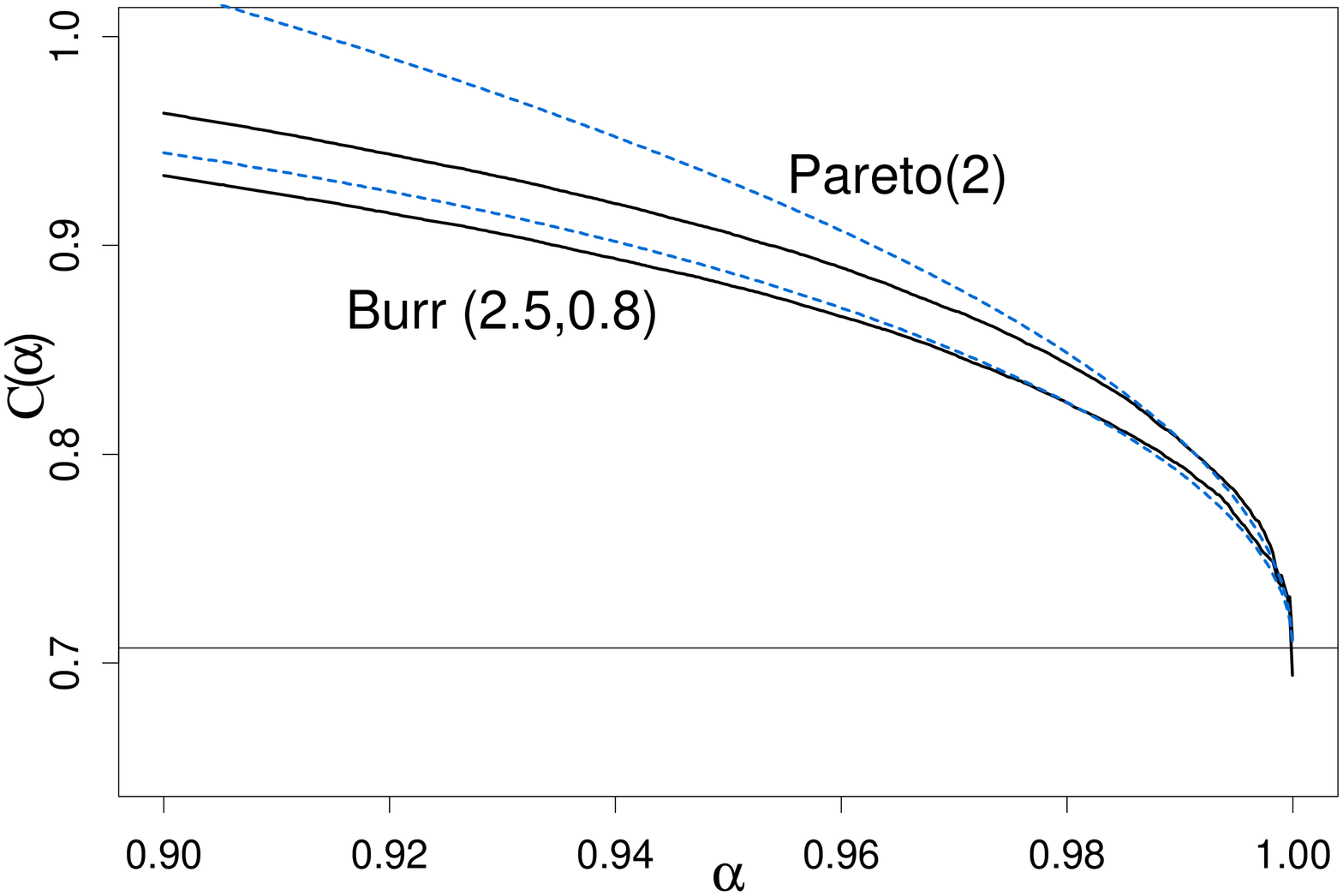, width = 0.49\textwidth}
\hspace{0mm}
      \epsfig{angle=-90,file=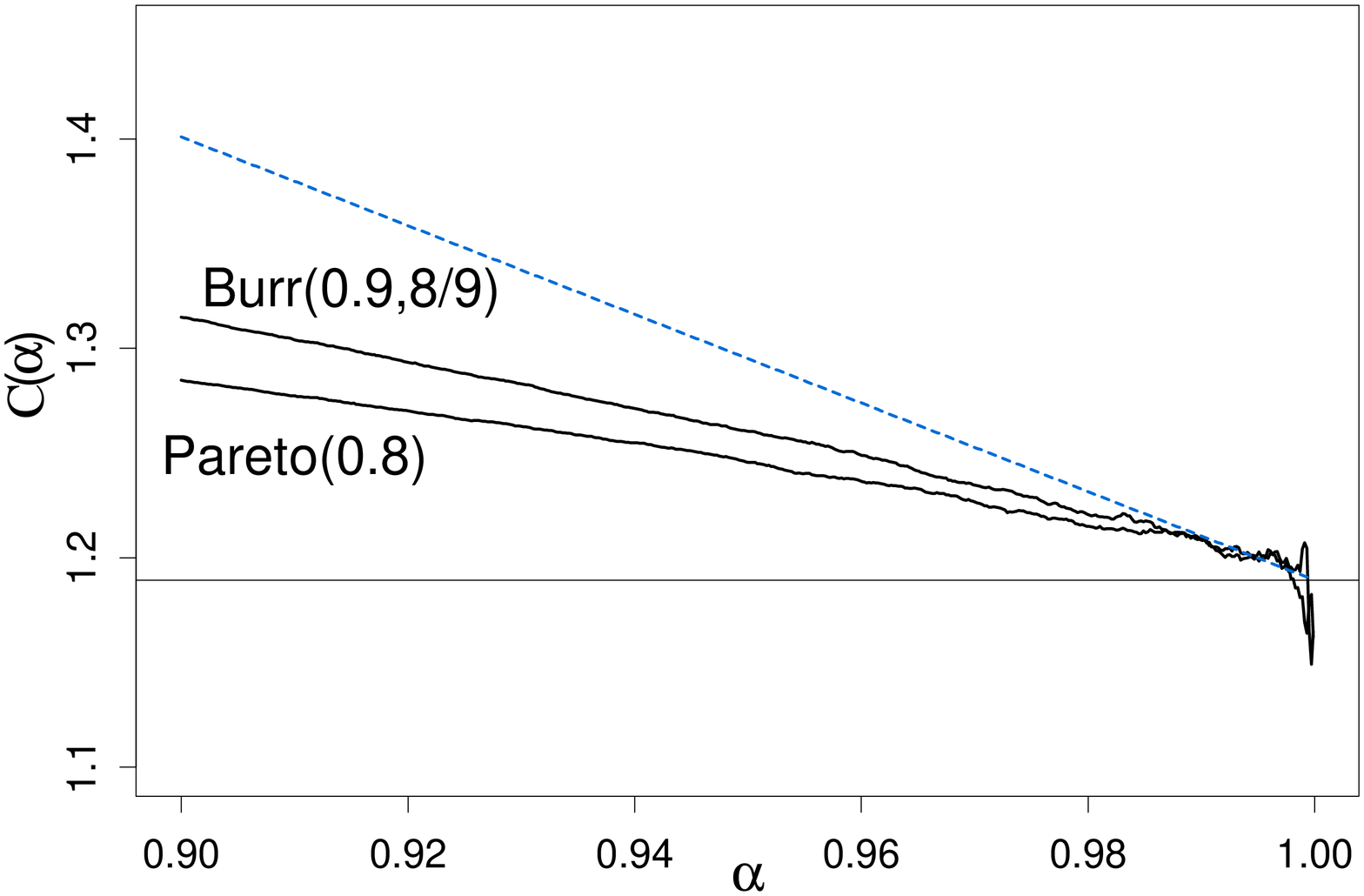, width = 0.49\textwidth}
\end{center}
\vspace{-1cm}
\caption{Empirical risk concentration (full, based on $10^7$ simulations) together with first-order approximation $C_1 \equiv 1/\sqrt{2} \approx 0.71$ (left panel) and $C_1\equiv2^{1/4}\approx 1.19$ (right panel) and second-order approximation $C_2$ (dashed) for two iid Burr($\tau,\kappa$) and Pareto($1/\xi$) random variables for a finite mean case (left panel) and an infinite mean case (right panel). Note the different scales on the vertical axis.}
   \label{fig:burr}
\end{figure}

\noindent
The gain of a second-order approximation $C_2$ over a first-order approximation $C_1$ is illustrated in Figure \ref{fig:burr} for the case of a fast converging Burr and an exact Pareto model.

For practical purposes, it is essential to know whether $C(\alpha)$ approaches its ultimate value $n^{\xi-1}$ from above or from below as $\alpha$ tends to 1. For a loss model $U \in 2RV_{\xi, \rho}$ satisfying the assumptions of Theorem \ref{thm1} with $\rho < -(1 \wedge \xi)$ (fast convergence case), the derivative of $C_2$ satisfies
\bd
\lim_{\alpha \rightarrow 1}C_2'(\alpha) \; = \; \begin{cases}
-\infty, &\textrm{if}\; \xi < 1,\; \textrm{or}\; \xi=1,\mu_F = \infty,\\
-\frac{n-1}{n}\frac{\mu_F}{c}, &\textrm{if}\; \xi = 1,\mu_F < \infty,\\
n^{\xi-2}(n-1)\xi\frac{\Gamma^2(1-1/\xi)}{2\Gamma(1-2/\xi)}, &\textrm{if}\; \xi > 1,
\end{cases}
\ed
for some $\displaystyle{c = \lim_{t \rightarrow \infty}L_U(t) \in (0, \infty)}$. Moreover, we have $C_2'(1) > 0$ if and only if $\xi > 2$, i.e.~it is only in very heavy-tailed cases that $C(\alpha)$ \paars{increases to} $n^{\xi-1}$ as $\alpha \rightarrow 1$.

In the case of \paars{slowly} converging loss models, the situation is more involved. For $U \in 2RV_{\xi,\rho}$ with $\rho > -(1 \wedge \xi)$, the ultimate behavior of $C_2'$ depends on the exact form of $L_U$ so that in general no precise statement can be made without extra assumptions. \paars{Still, for distributions in the Hall class, $C(\alpha)$ will approach its limit $n^{\xi-1}$ from above (below) if $d$ is negative (positive).}

Within the class of \paars{slowly} converging loss models, the case $\rho = 0$ deserves special attention, as in this case the decay of $|a|$ in (\ref{2RV}) may be arbitrarily slow. This is due to the behavior of the associated slowly varying function $L_U$ which, in the case $\rho = 0$, may indeed be rather misleading; see for instance \cite{DE}, Section~4. A prime example for this is provided for instance by Tukey's g-and-h distribution.

\begin{bsp}[\textnormal{g-and-h}]
\label{ex:gandh}
A random variable $X$ is said to follow Tukey's g-and-h distribution with parameters $a,b,g,h \in \mathds{R}$, if $X$ satisfies
\bd
X = a + b \frac{e^{gZ}-1}{g} e^{hZ^2/2},
\ed
where $Z \sim \mathcal{N}(0,1)$ and with the obvious interpretation for $g = 0$. Note that in principle such random variables need not be positive. In for financial risk management practice relevant cases the parameters typically satisfy $b,g, h> 0$, so that we may bypass this issue using the notion of {\it right tail dominance}; see \cite{Barbe}.

Suppose $X_1, \ldots, X_n$ are iid g-and-h random variables with $a=0$, $b=1$ and $g, h > 0$. Then one shows that $U \in 2RV_{\xi,\rho}$ with $\xi = h$, $\rho=0$ and that
\begin{eqnarray*}
a\left(\frac{1}{1-\alpha}\right) &=& \frac{U'\left(\frac{1}{1-\alpha}\right)}{(1-\alpha)U\left(\frac{1}{1-\alpha}\right)} - \xi\\
&=& \frac{g}{\Phi^{-1}(\alpha)} + O\left( \frac{1}{(\Phi^{-1}(\alpha))^2} \right), \quad \alpha \rightarrow 1,
\end{eqnarray*}
where we used the standard asymptotic expansion for the tail of the normal distribution given by $\overline{\Phi}(x) =e^{-x^2/2}/(\sqrt{2\pi}x)\left(1 + O\left(1/x^2\right)\right)$, $x \rightarrow \infty$; see for instance \cite{ABR-STEGUN}, p.~932. Therefore, we obtain the following second-order asymptotics for the risk concentration:
\bd
C(\alpha) = n^{h -1} +n^{h -1} \log (n)  \frac{g}{\Phi^{-1}(\alpha)} + o\left( \frac{1}{\Phi^{-1}(\alpha)}\right), \quad \alpha \rightarrow 1.
\ed
\end{bsp}

\begin{figure}[htb!]
\begin{center}
\setcaptionwidth{\textwidth}
   \psfig{file=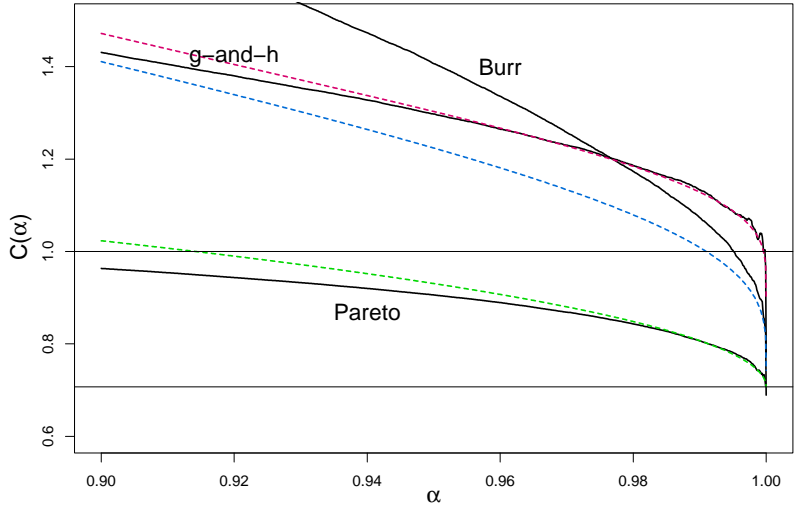,width=0.6\textwidth, angle=-90}
   \vspace{-0.5cm}
   \caption{Empirical risk concentration (full, based on $n=10^7$ simulations) together with \paars{the} first-order approximation $C_1(\alpha) \equiv 1/\sqrt{2} \approx 0.71$ and \paars{the} second-order approximation $C_2$ (dashed) for two iid Burr ($\tau = 0.25, \kappa = 8$), Pareto ($\xi = 0.5$) and g-and-h ($g=2, h=0.5$) random variables.}
   \label{fig:comparison-sec}
  \end{center}
\end{figure}

\noindent
Depending on the parameter values of $g$ and $h$, $C(\cdot)$ may be growing extremely fast when moving away from $\alpha =1$; see Figure \ref{fig:comparison-sec}. In that figure we compare a g-and-h with a Burr model of slow convergence and with a standard Pareto model. Note that we choose the \paars{tail index} as $\xi = 0.5$ in each model.

Figure \ref{fig:comparison-sec} allows us to draw the following conclusions. \paars{Even at high} levels of $\alpha<1$, the diversification benefit \paars{promised by first-order theory} may vanish rather quickly and may even get negative. For the g-and-h model of Figure \ref{fig:comparison-sec}, the regime switch from sub- to superadditivity takes place at the extreme level of $\alpha \approx 99.95\%$. \paars{The} second-order approximation $C_2$ is able to capture this behavior better than $C_1$.

\section*{Appendix}
\begin{Proof}[Theorem \ref{thm1}]
For $\alpha < 1$, define $x = G^{\leftarrow}(\alpha)=(F^{n\ast})^{\leftarrow}(\alpha)$. Note that the convergence in the \paars{definition} of second-order regular variation holds locally uniformly on $(0,\infty)$; see \cite{deHaan}, Remark B.3.8. Therefore, replacing $t$ by $1/\overline{G}(x)$ and $s$ by $\overline{G}(x)/\overline{F}(x)$ in (\ref{2RV}), $U \in 2RV_{\xi,\rho}$ implies
\bd
\lim_{x \rightarrow \infty} \frac{\frac{U(1/\overline{F}(x))}{nU(1/\overline{G}(x))}-\frac{1}{n}\left( \frac{\overline{G}(x)}{\overline{F}(x)} \right)^\xi}{a(1/\overline{G}(x))}
= n^{\xi-1} \frac{n^{\rho}-1}{\rho} = n^{-1}H_{\xi,\rho}(n),
\ed
with the obvious interpretation for $\rho=0$. From Proposition \ref{prop1} we then get
\begin{eqnarray*}
C(\alpha) &=& \frac{1}{n}\left( \frac{\overline{G}(x)}{\overline{F}(x)} \right)^\xi + n^{-1}H_{\xi,\rho}(n)a\left(\frac{1}{1-\alpha}\right)
+ o\left(a\left(\frac{1}{1-\alpha}\right)\right)\\
&=& n^{\xi-1} (1 + \xi n^{-1} J_{\xi}(n) b(G^{\leftarrow}(\alpha))) + n^{-1} H_{\xi,\rho}(n)a\left(\frac{1}{1-\alpha}\right)\\
&&+ o\left( b(G^{\leftarrow}(\alpha)) \right) + o\left(a\left(\frac{1}{1-\alpha}\right)\right),
\end{eqnarray*}
for $\alpha \rightarrow 1$ and where we have used the expansion $(1 + y)^{\xi} = 1 + \xi y + o(y)$ as $y \rightarrow 0$. Note that, due to regular variation of $b$, we have
\bd
\frac{b(G^\leftarrow(\alpha))}{b(F^\leftarrow(\alpha))} \sim \left( \frac{G^\leftarrow(\alpha)}{F^\leftarrow(\alpha)} \right)^{-(1 \wedge 1/\xi)}
\rightarrow n^{-(\xi \wedge 1)}, \quad  \alpha \rightarrow 1.
\ed
Define $\widetilde{A}(\alpha) = b(F^\leftarrow(\alpha)) + a\left(\frac{1}{1-\alpha}\right)$. Note that $b\circ F^{\leftarrow} \in RV_{-(1 \wedge \xi)}$ and $|a| \in RV_\rho$. Due to the regular variation properties of $b$ and $a$ and since $\rho \neq -(1\wedge \xi)$ this implies that
\bd
\frac{C(\alpha) - n^{\xi-1}}{\widetilde{A}(\alpha)} = \xi n^{\xi-2} n^{-(\xi \wedge 1)} J_{\xi}(n) \frac{b(F^\leftarrow(\alpha))}{\widetilde{A}(\alpha)}
+ n^{-1}H_{\xi,\rho}(n) \frac{a\left(\frac{1}{1-\alpha}\right)}{\widetilde{A}(\alpha)} +o(1),
\ed
as $\alpha \rightarrow 1$, which yields the result.
\end{Proof}

\begin{Proof}[Remark \ref{rmk:aux-function}]

$U \in 2RV_{\xi, \rho}(a)$ (with $\rho \le 0 < \xi$) with auxiliary function $a(\cdot)$ implies $U \in RV_{\xi}$ and we write $U(t) = t^{\xi}L(t)$ for some slowly varying function $L$. With this notation and for $s > 0$,
\bd
\lim_{t \rightarrow \infty}\frac{\frac{U(ts)}{U(t)} - s^{\xi}}{a(t)} = s^{\xi}\frac{s^{\rho} - 1}{\rho} \Longleftrightarrow \lim_{t \rightarrow \infty}\frac{L(ts) - L(t)}{a(t)L(t)} = \frac{s^{\rho} - 1}{\rho}.
\ed
Hence $L \in ERV_{\rho}(B)$, i.e.~L is extended regularly varying with index $\rho \leq 0$ and auxiliary function $B(t) = a(t)L(t)$. For an introduction to ERV we refer to \cite{deHaan}, Appendix B.2.

Case $\rho = 0$: We write $L(t) = L(t_0) + \int_{t_0}^tL'(s)ds$. The ultimate monotonicity of $L'$ guarantees $L' \in RV_{-1}$ by the Monotone Density Theorem for $\Pi$-variation; see \cite{BINGHAM}, Corollary 3.6.9. In that case, $tL'(t)$ is an auxiliary function, hence necessarily $B(t) \sim tL'(t), t \rightarrow \infty$; see \cite{deHaan}, Remark B.2.6.

Case $\rho < 0$: In that case, the limit $\lim_{t \rightarrow \infty}L(t) = L(\infty)$ exists and is finite. Set $f(t) = L(\infty) - L(t) = \int_{t}^{\infty}L'(s)ds$. Then $\lim_{t\rightarrow \infty}\frac{f(t)}{B(t)} = -1/\rho$ and $f(t) \in RV_{\rho}$, by Theorem B.2.2 in \cite{deHaan}. Ultimate monotonicity of $L'$ implies that $\frac{tL'(t)}{f(t)} \rightarrow -\rho$ by Proposition B.1.9 11) of \cite{deHaan} and hence $B(t) \sim tL'(t)$ as $t \rightarrow \infty$.

Altogether, we thus obtain $a(t) = \frac{B(t)}{L(t)} \sim \frac{tL'(t)}{L(t)} = \frac{tU'(t)}{U(t)} - \xi$ as $t \rightarrow \infty$.
\end{Proof}

%%%%%%%%%%%%%%%%%%%%%%%%%%%%%%%%%%%%%%%%%%%%%%%%%%%%
%%%%%%%%%%%%%%%%%%%%%%%%%%%%%%%%%%%%%%%%%%%%%%%%%%%%
%%%%%%%%%%%%%%%%%%%%%%%%%%%%%%%%%%%%%%%%%%%%%%%%%%%%
%%%%%%%%%%%%%%%%%%%%%%%%%%%%%%%%%%%%%%%%%%%%%%%%%%%%

\section*{Acknowledgments}
The authors would like to thank Paul Embrechts and Mario V.~W\"uthrich for several discussions related to this paper and the numerous useful comments as well as the referee for constructive suggestions. The first two authors gratefully acknowledge the financial support by RiskLab Switzerland. The research of the third author is supported by IAP research network grant nr.\ P6/03 of the Belgian government (Belgian Science Policy) and by contract nr.\ 07/12/002 of the Projet d'Actions de Recherche Concert\'ees of the Communaut\'e fran\c{c}aise de Belgique, granted by the Acad\'emie universitaire Louvain.

%\section*{References}
\bibliography{bibliography}
\bibliographystyle{elsarticle-harv}
\end{document}